\documentclass[sigconf]{acmart}
\AtBeginDocument{%
  }

\setcopyright{acmlicensed}
\copyrightyear{2018}
\acmYear{2018}
\acmDOI{XXXXXXX.XXXXXXX}
\acmConference[Conference acronym 'XX]{Make sure to enter the correct
  conference title from your rights confirmation email}{June 03--05,
  2018}{Woodstock, NY}
\acmISBN{978-1-4503-XXXX-X/2018/06}

\usepackage{xurl} 
\usepackage{multirow}
\usepackage{enumitem}
\usepackage[utf8]{inputenc}
\usepackage[most]{tcolorbox} 
\usepackage{amsmath}         

\newtcolorbox{promptbox}[1]{
    colback=gray!10,           
    colframe=black,            
    colbacktitle=black,        
    coltitle=white,            
    fonttitle=\bfseries\large, 
    title={#1},                
    arc=10pt,                  
    outer arc=10pt,
    left=10pt,                 
    right=10pt,                
    top=10pt,                  
    bottom=10pt,               
    boxrule=1.5pt,             
    enhanced                   
}

\begin{document}

\title{AIGQ: An End-to-End Hybrid Generative Architecture for E-commerce Query Recommendation}
\author{Jingcao Xu}
\authornote{Equal contribution.}
\authornote{Corresponding author.}
\email{xujingcao.xjc@taobao.com}
\affiliation{%
  \institution{Taobao \& Tmall Group of Alibaba}
  \city{Hangzhou}
  \country{China}
}

\author{Jianyun Zou}
\authornotemark[1]
\email{zoujianyun.zjy@taobao.com}
\affiliation{%
  \institution{Taobao \& Tmall Group of Alibaba}
  \city{Hangzhou}
  \country{China}
}

\author{Renkai Yang}
\authornotemark[1]
\email{renkai.yrk@taobao.com}
\affiliation{%
  \institution{Taobao \& Tmall Group of Alibaba}
  \city{Hangzhou}
  \country{China}
}

\author{Zili Geng}
\email{gengzili@std.uestc.edu.cn}
\affiliation{%
  \institution{University of Electronic Science and Technology of China}
  \city{Chengdu}
  \country{China}
}

\author{Qiang Liu}
\email{2012dtd@gmail.com}
\affiliation{%
  \institution{Taobao \& Tmall Group of Alibaba}
  \city{Hangzhou}
  \country{China}
}

\author{Haihong Tang}
\email{piaoxue@taobao.com}
\affiliation{%
  \institution{Taobao \& Tmall Group of Alibaba}
  \city{Hangzhou}
  \country{China}
}



\begin{abstract}

Pre-search query recommendation, widely known as HintQ on Taobao’s homepage, plays a vital role in intent capture and demand discovery, yet traditional methods suffer from shallow semantics, poor cold-start performance and low serendipity due to reliance on ID-based matching and co-click heuristics. To overcome these challenges, we propose AIGQ (AI-Generated Query architecture), the first end-to-end generative framework for HintQ scenario. AIGQ is built upon three core innovations spanning training paradigm, policy optimization and deployment architecture. First, we propose Interest-Aware List Supervised Fine-Tuning (IL-SFT), a list-level supervised learning approach that constructs training samples through session-aware behavior aggregation and interest-guided re-ranking strategy to faithfully model nuanced user intent. Accordingly, we design Interest-aware List Group Relative Policy Optimization (IL-GRPO), a novel policy gradient algorithm with a dual-component reward mechanism that jointly optimizes individual query relevance and global list properties, enhanced by a model-based reward from the online click-through rate (CTR) ranking model. To deploy under strict real-time and low-latency requirements, we further develop a hybrid offline-online architecture comprising \textsc{AIGQ-Direct} for nearline personalized user-to-query generation and \textsc{AIGQ-Think}, a reasoning-enhanced variant that produces trigger-to-query mappings to enrich interest diversity. Extensive offline evaluations and large-scale online A/B experiments on Taobao demonstrate that AIGQ consistently delivers substantial improvements in key business metrics across platform effectiveness and user engagement.

\end{abstract}


\maketitle

\section{Introduction}
\begin{figure}[htbp]
    \centering
    \includegraphics[width=\linewidth]{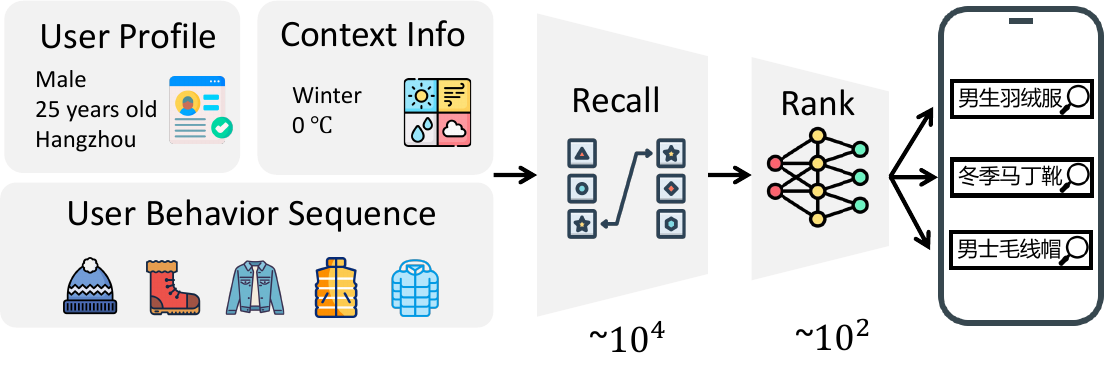}
    \caption{Overview of HintQ}
    \label{fig:preQ}
\end{figure}

In modern search engines, query recommendation systems are routinely deployed to assist users in articulating their information needs. These systems typically fall into three canonical categories: (1) Query Suggestion\cite{li2014two, zamani2020generating, zhang2015adaqac}, which predicts query suffixes or full queries based on a user-provided prefix; (2) Query Clarification\cite{10.1145/3397271.3401160,hashemi2020guided,pyatkin2023clarifydelphi,zamani2020analyzing}, which resolves query ambiguity through disambiguating prompts; (3) Query Recommendation\cite{min2025promoting, min-etal-2025-ctr, lee2024enhanced,li2019click}, which proposes alternative queries derived from aggregated behavioral data or trending signals. Within this category, HintQ is a pre-search query recommendation feature on Taobao’s homepage that operates without any current query context, taking the user’s historical behavior sequence and profile attributes as input and generating personalized queries as output, as illustrated in Figure~\ref{fig:preQ}.

Early query recommendation systems relied on collaborative filtering\cite{5197422} and DNN-based matching \cite{10.1145/3038912.3052569, 10.1145/2959100.2959190} which suffer from poor cold-start generalization, but the recent rise of Large Language Models (LLMs) has sparked a paradigm shift toward generative recommendation, enabling diverse applications such as item recommendation~\cite{rajput2023recommender,liang2025tbgrecall,xing2025reg4rec} and video suggestion\cite{zhou2025onerec, zhou2025onerecv2, liu2025onerec}. However, no prior work has explored generative ability for hint query recommendation. While OneSug\cite{guo2025onesug} represents a pioneering effort in LLM-based query recommendation, it optimizes primarily for relevance under strong contextual constraints. In contrast, HintQ must generate diverse, personalized queries without any trigger, which poses a great challenge for requiring a delicate balance between personalization accuracy and serendipitous discovery.

In industrial practice, the common approach to leveraging LLMs is to construct a static context-to-query mapping offline and then function as an online recall in response to user triggers. However, this paradigm fails to capture the multi-faceted nature of user interests and often produces incoherent outputs. Moreover, these queries are fed into a downstream ranker that optimizes short-term engagement (e.g., CTR), creating a misalignment between semantic legitimacy and user preference. To enable fully personalized query recommendation via LLMs in HintQ where multiple candidates are presented in ranked order, a key challenge lies in designing an end-to-end generative framework that directly optimizes the ranking of generated queries aligned with real user preferences.

As for the online deployment of generative architectures in industrial recommendation scenarios, existing end-to-end approaches treat recommendation as a next token prediction (NTP) task over semantic IDs and rely on online beam search to generate multiple candidates ~\cite{rajput2023recommender,yi2025dualgr,hao2025oxygenrec}. However, in industrial recommendation systems, it remains tempered in practice by tight constraints on computational resources and real-time latency budgets. Therefore, another key challenge arises in the balance of maximizing the semantic fidelity of LLM-based modeling while simultaneously achieving inference efficiency at scale.

To address these challenges, we propose AIGQ (AI-Generated Query architecture), a hybrid end-to-end generative framework for pre-search query recommendation in e-commerce. Through co-design of list-level training objectives, policy optimization and deployment strategy, AIGQ unifies semantic capability with inference efficiency, effectively balancing accuracy, diversity and latency under real-world constraints. To the best of our knowledge, AIGQ is the first end-to-end generative solution successfully deployed for the HintQ scenario, delivering significant improvements in both user engagement and business outcomes. The main contributions of this work are summarized as follows:
\begin{itemize}[left=0pt]
    \item We present the first end-to-end generative framework for pre-search query recommendation scenario (HintQ), introducing two specialized LLM variants, i.e., \textsc{AIGQ-Direct} and \textsc{AIGQ-Think}, designed to address the dual demands of high accuracy and broad interest coverage. Both models are trained on massive user behavior logs using an innovative \textit{Interest-Aware List Supervised Fine-Tuning} (\textbf{IL-SFT}) paradigm that jointly optimizes personalization fidelity and inference efficiency.
    \item We propose \textit{Interest-aware List Group Relative Policy Optimization} (\textbf{IL-GRPO}), a novel policy gradient algorithm that performs advantage estimation at the granularity of individual queries within a generated list. IL-GRPO employs a dual-component reward mechanism, combining a local query-level reward that assesses the quality and relevance of individual recommended queries with a global sequence-level reward that evaluates the overall coherence, coverage and diversity of the entire recommendation list.
    \item We design a hybrid offline-online deployment architecture for end-to-end generative models, where lightweight \textsc{AIGQ-Direct} generates nearline personalized user-to-query (u2q) hints and \textsc{AIGQ-Think} constructs offline trigger-to-query (x2q) mappings to support real-time refinement. The system meets stringent online latency requirements while delivering remarkable metrics improvement and the highest PVR contribution, demonstrating the feasibility of \textsc{AIGQ} in production.
\end{itemize}


\section{Related Work}

\subsection{Query Recommendation} 
Traditional query recommendation methods use heuristic rules and statistical models based on query logs, term co-occurrence, and user behavior. They work well in narrow settings but often suffer from sparse data and struggle to adapt to changing user intents~\cite{10.1007/978-3-540-30192-9_58, 10.1145/2806416.2806493}. Neural approaches improve this by modeling sequential patterns for better contextual recommendations~\cite{lai2023workload}, yet they still struggle with long-tail queries. Recently, the field has moved toward generative methods using LLMs. For example, GQR~\cite{bacciu2024generating} handles cold-start cases without historical logs. Later work boosts relevance by linking generation to CTR prediction~\cite{min2025promoting} or using query-to-recommendation frameworks for personalized expansion~\cite{han2025query2rec}. Other studies show that LLM-generated pseudo references greatly improve retrieval~\cite{zhang2024mugi}, and tailored techniques can produce high-quality queries even in low-resource, task-oriented systems~\cite{chen2025llmenhanced}.

\subsection{Generative Recommendation} 
Generative recommendation leverages LLMs to produce recommendations as sequences, either in the form of semantic IDs (SIDs) or natural language. For instance, TIGER~\cite{rajput2023recommender} generates SIDs via a sequence-to-sequence framework, while HLLM~\cite{chen2024hllm} employs a hierarchical structure to model user interests and item representations separately. In industrial practice, the OneRec family~\cite{zhou2025onerec, zhou2025onerecv2, liu2025onerec} establishes end-to-end generative frameworks that integrate reasoning and reinforcement learning for video and e-commerce recommendations. Concurrently, SynerGen~\cite{gao2025synergen} and OneSearch~\cite{chen2025onesearch, guo2025onesug} unify search and recommendation tasks under generative paradigms. However, recent work suggests natural language itself can serve as a more expressive alternative to SIDs~\cite{liu2025understanding}. Distinct from SID-based approaches, \textsc{AIGQ} operates entirely in natural language space, directly generating contextually persuasive queries without discrete indexing, thereby fully harnessing the generative capacity of LLMs for pre-search query recommendation.


\subsection{Preference Alignment for Recommendation}
Reinforcement learning (RL) has emerged as a powerful approach to align generative recommenders with real user preferences beyond static supervision. In this direction, Rec-R1~\cite{lin2025recr1bridginggenerativelarge} shows that closed-loop policy optimization adapts LLM-based generation more effectively than conventional prompting or fine-tuning, while STREAM-Rec~\cite{zhang2025slowthinkingsequentialrecommendation} applies RL to improve deliberative reasoning in sequential recommendation tasks. Industrial systems further demonstrate the value of reinforcement learning, with OneRec~\cite{zhou2025onerec} employing multi-dimensional rewards to balance business constraints and user engagement, and OneRec-Think~\cite{liu2025onerec} using RL to refine explicit chain-of-thought (CoT) reasoning for complex user intents. However, applying standard RL to list-wise query generation remains challenging due to high computational cost and strong inter-query dependencies. To address this, we adapt GRPO~\cite{shao2024deepseekmathpushinglimitsmathematical} to our setting by introducing fine-grained advantage estimation that operates at the granularity of individual queries within a generated list.

\section{Methodology}
\begin{figure*}[t]
    \centering
    \setlength{\abovecaptionskip}{-0.5em}
    \setlength{\belowcaptionskip}{-1em}
    \includegraphics[width=1\linewidth]{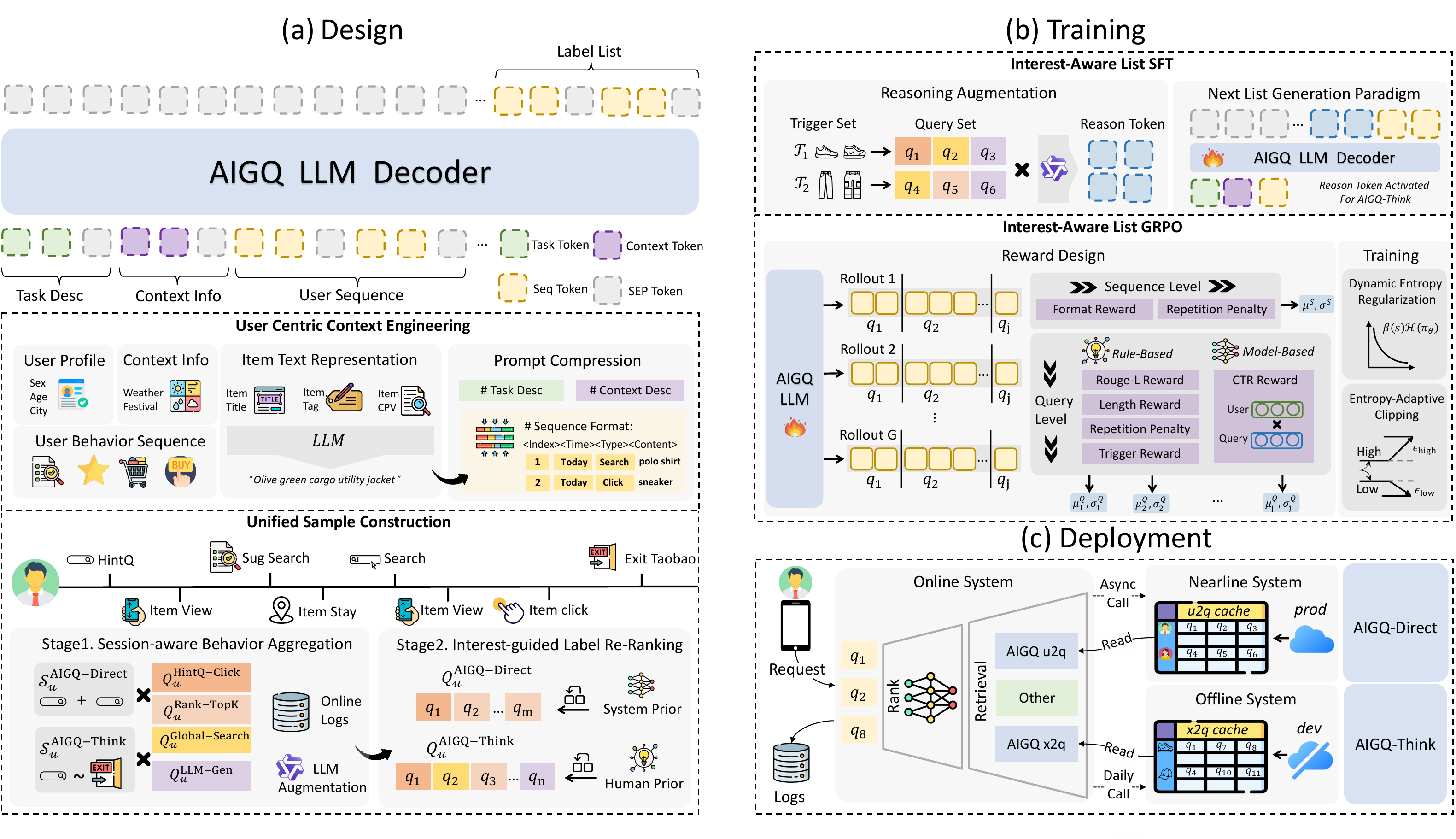}
    \caption{Overview of the \textsc{AIGQ} framework.}
    \label{fig:framework}
\end{figure*}

In this section, we elaborate on our AIGQ framework in detail. We first define the problem for HintQ recommendation task, then we present our design in context engineering and sample construction, followed by the introduction of the two-stage training paradigm. The overall framework is depicted in Figure \ref{fig:framework}.
\subsection{Problem Definition}
\label{sec:problem}

In this work, we focus on the HintQ recommendation task in Taobao’s search scenario. Formally, given a user $u$ with demographic profile $\mathcal{P}_u$ (e.g., age, gender, location) and a time-ordered behavior sequence $\mathcal{S}_u = \{(t_i, a_i, c_i)\}_{i=1}^L$, where $t_i$ denotes the time offset (in days ago), $a_i \in \{\text{search}, \text{click}\}$ indicates the action type, and $c_i$ is the associated content (query text or item description), the goal is to generate a ranked list of $K$ hint queries $\mathcal{Q} = [q_1, q_2, \dots, q_K]$. This list must capture the user’s near-term intent, jointly optimize for relevance and business utility (e.g., CTR, GMV).

Importantly, in contrast to conventional two-stage pipelines, which first retrieve candidates via pointwise scoring and subsequently rerank them, we formulate HintQ recommendation as an interest-aware list generation task, which produces the entire ordered candidate list in a holistic manner to eliminate the need for online beam search, thereby reducing both computational overhead and inference latency. 

\subsection{User-Centric Context Engineering}

In personalized query recommendation, the input representation plays a pivotal role in accurately capturing user intent and generating relevant queries. Given the sensitivity of LLMs to input length and inference costs, we adopt a deliberately minimal user-level context: basic user profile attributes (e.g., age, gender, region) are verbalized in natural language, and user behavior histories are merged into a time-ordered sequence using a unified textual schema \textit{"index | time | behavior type | behavior content"}. The remainder of this section focuses on two key components: behavior content representation and prompt compression.

\subsubsection{Behavior Content Representation.} In this work, we model user behavior history using search queries and item clicks. Search queries are naturally textual, whereas items are inherently multimodal, encompassing both rich textual attributes (e.g., titles, tags) and visual content (e.g., product images). To bridge the semantic gap between discrete item identifiers and the continuous representation space of LLMs, we design an item-to-text generator that synthesizes textual surrogates from structured item metadata. Specifically, we fine-tune a Qwen3-32B~\cite{yang2025qwen3technicalreport} language model on large-scale user click logs to generate concise descriptions from heterogeneous item facets including title, category, attributes, tags, etc., which encapsulate item’s core characteristics and user-preferred linguistic patterns.

\subsubsection{Prompt Compression.}
To minimize token consumption while preserving context integrity, we compress both feature representations and task prompts through two complementary strategies: (1) \textbf{Feature structuring.} Discrete fields with small vocabularies, such as timestamps and behavior types, are encoded as dedicated special tokens added to the model’s vocabulary, replacing verbose natural-language descriptions of the schema.

{\footnotesize
\begin{equation}
\text{``1|1 day ago\,|\,search\,|\,iPhone17''}
\Rightarrow
\texttt{<1><1\_day\_ago><search> iPhone17}.
\end{equation}
}

where the special token is initialized as the average of the embeddings of its constituent subwords, for example, the embedding of <1\_day\_ago> is initialized as the mean of the embeddings of "1", "day", and "ago". This substantially shortens long behavior sequences while keeping their structure explicit. (2) \textbf{ Instruction pruning.} Leveraging the fact that supervised fine-tuning (SFT) has already internalized the task objective and input schema, we adopt a minimal instruction format and eliminate redundant explanatory text in the final prompt. The complete prompt example is as follows: 

\begin{promptbox}{AIGQ Prompt.}\small
    \textbf{Input:} Please Recommend the next 10 queries. \\
    User Profile: \{ \textit{Gender: Female, Age: 44, Occupation: Medical Staff, City: Beijing} \}. \\
    User Behavior Sequence: \{ \textit{<1><Today><Click> National Style Hand-woven Sachet;; 
    <1><Today><Search> Flower Cake Sachet...} \}
    \medskip \\
    \textbf{Output:} \{Predicted queries.\}
\end{promptbox}


\subsection{Unified Sample Construction Framework}
\label{subsec:sample_construction}

To support both domain-specific precision and global interest exploration, we adopt a unified two-stage sample construction pipeline: (1) \textbf{session-aware behavior aggregation}, which defines the temporal scope of user intent and gathers raw interaction signals; and (2) \textbf{interest-guided label re-ranking}, which organizes the candidate queries into an ordered supervision list according to quantified interest strength. While sharing this structural backbone, \textsc{AIGQ-Direct} and \textsc{AIGQ-Think} differ significantly in their session definitions and label curation strategies, as detailed below.

\subsubsection{Stage 1: Session-Aware Behavior Aggregation}
The session boundary governs the granularity of intent modeling and determines which interactions constitute a valid training instance. Let $\mathcal{H}_u$ denote the complete history of user $u$’s timestamped interactions, including query exposures, clicks, and other in-app actions.

In \textsc{AIGQ-Direct}, a session $\mathcal{S}_u^{\textsc{AIGQ-Direct}}$ is defined as a single exposure event in the HintQ scenario, where three hint queries are presented to the user in one request. This exposure is retained as a valid session only if at least one of the three displayed queries receives a click,  as verified using the user’s click records in $\mathcal{H}_u$. This fine-grained definition ensures high behavioral fidelity but restricts the context to a narrow in-domain snapshot. 

In contrast, \textsc{AIGQ-Think} adopts a day-level session definition:

{\footnotesize
\begin{equation}
\mathcal{S}_u^{\textsc{AIGQ-Think}} = \{ h \in \mathcal{H}_u \mid t_{\text{first}} \leq h.\text{time} \leq t_{\text{exit}} \}
\end{equation}
}

where $t_{\text{first}}$ marks the user’s first interaction with the search entry and $t_{\text{exit}}$ denotes their final exit from Taobao on the same day. This window aggregates cross-domain behaviors which enables the discovery of diverse interests beyond immediate exposure.


\subsubsection{Stage 2: Interest-Guided Label Re-Ranking}
Given a valid session, we construct a ranked label list $\mathcal{Q}_u = [q_1, \dots, q_K]$ by reorganizing the collected candidates according to interest strength derived from system priors (e.g., production ranking scores) and feedback priors (e.g., clicks outweighing exposures), which serves as the supervision target for list generation fine-tuning.

Intuitively, through years of optimization based upon the real-world engagement feedback, the production ranking model is able to encode user interest strength implicitly. Therefore, for \textsc{AIGQ-Direct}, the labels are composed of the clicked hint query $\mathcal{Q}_u^{\text{HintQ-Click}}$ and top ranked queries $\mathcal{Q}_u^{\text{Rank-TopK}}$ during the current exposure event. Finally, the ranked label list is as follows:

{\footnotesize
\begin{equation}
\mathcal{Q}_u^{\textsc{AIGQ-Direct}} = \texttt{Concat(}\mathcal{Q}_u^{\text{HintQ-Click}}, \mathcal{Q}_u^{\text{Rank-TopK}}) 
\end{equation}
}

In contrast, \textsc{AIGQ-Think} explicitly models multi-faceted user interests by aggregating heterogeneous query signals from the user’s day-level session, including in-domain signals (i.e., the clicked hint query and top ranked queries), cross-domain global search queries, and LLM-generated candidates. Based on their reliability as behavioral indicators, we construct the ranked label list as follows:

{\footnotesize
\begin{equation}
\mathcal{Q}_u^{\textsc{AIGQ-Think}} = \texttt{Concat}\big( \mathcal{Q}_u^{\text{HintQ-Click}},\, \mathcal{Q}_u^{\text{Global-Search}},\, \mathcal{Q}_u^{\text{Rank-TopK}},\, \mathcal{Q}_u^{\text{LLM-Gen}} \big)
\end{equation}
}

where $\mathcal{Q}_u^{\text{HintQ-Click}}$ and $\mathcal{Q}_u^{\text{Global-Search}}$ represent high-confidence positive feedback signals, while $\mathcal{Q}_u^{\text{Rank-TopK}}$ and $\mathcal{Q}_u^{\text{LLM-Gen}}$ provide coverage expansion with decreasing reliability. To ensure label relevance to user behavior, we further employ a powerful LLM to filter out candidates that exhibit weak alignment with the user’s historical behavioral patterns.

\subsection{Interest-Aware List SFT (IL-SFT)}
\label{subsec:list_gen_paradigm}
In this section, we propose \textbf{Interest-Aware List SFT (IL-SFT)}, a supervised fine-tuning framework that explicitly models the target as an ordered list of queries grounded in quantifiable user interests. 
Critically, IL-SFT diverges from conventional next-token prediction (NTP) by treating the target not as an unstructured token sequence, but as an ordered interest-driven list. 
The training objective is formulated as:

{\footnotesize
\begin{equation}
\mathcal{L}_{\text{SFT}} = -\mathbb{E}_{(\mathbf{x}, \mathbf{z}) \sim \mathcal{D}} \left[ \sum_{t=1}^{T} \log P(z_t \mid z_{<t}, \mathbf{x}; \theta) \right],
\end{equation}
}

where $\mathbf{x} = [\mathcal{P}, \mathcal{S}]$ denotes the user context, $\theta$ denotes model parameters and $T$ is the output sequence length. The key distinction lies in the composition of the target sequence $\mathbf{z}$. Both \textsc{AIGQ-Direct} and \textsc{AIGQ-Think} instantiate this paradigm, differing only in the internal structure of the target sequence. 

\subsubsection{\textsc{AIGQ-Direct}} We generate a flat list of top-$K$ queries directly based on the output of the production ranking model. The training dataset is constructed as:

{\footnotesize
\begin{gather}
\mathcal{D}_{direct} = \left\{ \big( \mathbf{x}^{(n)},\,  \mathbf{y}^{(n)} \big) \right\}_{n=1}^N, \ \ \
\mathbf{y}^{(n)} = \mathcal{Q}^{(n)}
\end{gather}
}
where $N$ denotes the number of training samples, $\mathcal{Q}^{(n)}$ denote the label list of the $n$-th sample, which instantiates $\mathcal{Q}_u^{\textsc{AIGQ-Direct}}$.


\subsubsection{\textsc{AIGQ-Think}} Inspired by recent advances in chain-of-thought (CoT) reasoning for recommendation, we enhance \textsc{AIGQ-Think} with explicit reasoning augmentation, which follows a two-stage cognitive process: (1) induce multi-interest points represented as trigger behavior sets $\{ \mathcal{T}_k \}_{k=1}^K$; (2) generate a query list $\mathcal{Q}_k$ for each interest point. The final output takes the structured form $\mathbf{y} = \{ (\mathcal{T}_1 \rightarrow \mathcal{Q}_1), \dots, (\mathcal{T}_K \rightarrow \mathcal{Q}_K) \}$. To make this reasoning process explicit and distillable, we use Qwen3-32B as a teacher to generate natural-language rationales. Specifically, for each interest point $(\mathcal{T}_k, \mathcal{Q}_k)$, we prompt the teacher model to produce a sub-reasoning chain $\mathbf{r}_k$ explaining the mapping from behaviors to queries:

{\footnotesize
\begin{equation}
\mathbf{r}_k = \text{LLM}_{\text{reason}}\big( \mathbf{x}, \mathcal{T}_k, \mathcal{Q}_k; \texttt{prompt} \big)
\end{equation}
}

The final output is formed by concatenating the reasoning trace with the structured prediction. Specifically, the complete reasoning trace is composed as accordingly $\mathbf{r} = \mathbf{r}_1 \,\|\, \cdots \,\|\, \mathbf{r}_K$, and the distillation dataset is constructed as:

{\footnotesize
\begin{gather}
\mathcal{D}_{think} = \left\{ \big( \mathbf{x}^{(n)},\, \mathbf{r}^{(n)},\, \mathbf{y}^{(n)} \big) \right\}_{n=1}^N
\\
\mathbf{r}^{(n)} = \{ \mathbf{r}_k^{(n)} \}_{k=1}^{K_n}, \ \ \ \mathbf{y}^{(n)} = \{ (\mathcal{T}_k^{(n)} \rightarrow \mathcal{Q}_k^{(n)}) \}_{k=1}^{K_n}
\end{gather}
}

where ${K_n}$ denotes the number of interest points for the $n$-th sample. Through this design, \textsc{AIGQ-Think} extends the output space to jointly model structured predictions and their underlying reasoning traces, thereby guiding the prediction process toward outputs that are not only accurate but also interpretable and controllable.
\subsection{Interest-aware List GRPO (IL-GRPO)}

In this section, we introduce \textbf{Interest-aware List GRPO (IL-GRPO)}, an enhanced reinforcement learning algorithm specifically designed for ordered list generation. IL-GRPO enhances fine-grained optimization based on GRPO by combining query-level and sequence-level rewards, thereby facilitating precise credit assignment that reflects both user preferences and reasoning rationality.

\subsubsection{Dual-Level Reward Design}
\label{subsubsec:query_reward}
Conventional 
reinforcement learning methods are ill-suited for list generation, as they assign a single scalar reward to the entire sequence, leading to severe credit assignment issues: high-quality queries may be unfairly penalized by weaker ones in the same list. To overcome this, we propose a fine-grained optimization strategy that decouples individual query evaluation from the sequence level.

Specifically, given that the output of IL-SFT implies strict order, we adopt a position-wise matching strategy, i.e., the $j$-th generated query is directly aligned with the $j$-th reference query. Based on this structure, we define two distinct reward components consisting of a local query-level reward and a global sequence-level reward:

{\footnotesize
\begin{equation}
\mathcal{R}_{\text{query}}(q_{i,j}) = \sum_{m} w_m r_m(q_{i,j}), \quad \mathcal{R}_{\text{seq}}(o_i) = \mathcal{R}_{\text{fmt}}(o_i) + \mathcal{R}_{\text{rep}}(o_i)
\label{eq:reward_definitions}
\end{equation}
}

where $w_m$ denotes the weight for the $m$-th reward component, $i$ denotes the sample index and $\mathcal{R}_{\text{query}}$ integrates three aspects of query quality: (1) \textit{user preference alignment}, approximated through the production CTR model; (2) \textit{semantic fidelity}, measured by ROUGE-L\cite{lin-2004-rouge}, which suits search queries due to its sensitivity to lexical and sequential overlap; and (3) \textit{query consistency}, enforced via length constraints to avoid truncation or unnatural elongation. Meanwhile, $\mathcal{R}_{\text{seq}}$ captures global sequence objectives with $\mathcal{R}_{\text{fmt}}$ ensuring proper punctuation and structural consistency, while $\mathcal{R}_{\text{rep}}$ penalizes redundant or near-duplicate queries to promote diversity. Formally, 
Let $\{o_1, \dots, o_G\}$ denote a group of $G$ outputs sampled from the current policy, the advantage $A_{i,t}$ for each token $t$ is computed by summing the normalized advantages from both levels, i.e., for the specific query position and the global sequence.

{\footnotesize
\begin{gather}
A_{i,t} = \underbrace{\frac{\mathcal{R}_{\text{query}}(q_{i,j}) - \mu_j^Q}{\sigma_j^Q}}_{\text{Query Advantage}} + \lambda \underbrace{\frac{\mathcal{R}_{\text{seq}}(o_i) - \mu^S}{\sigma^S}}_{\text{Sequence Advantage}},
\label{eq:fine_grained_advantage}
\\
\mu_j^Q = \frac{1}{G}\sum_{i=1}^{G} \mathcal{R}_{\text{query}}(q_{i,j}), \quad \sigma_j^Q = \sqrt{\frac{1}{G}\sum_{i=1}^{G} (\mathcal{R}_{\text{query}}(q_{i,j}) - \mu_j^Q)^2 + \epsilon}. 
\\
\mu^S = \frac{1}{G}\sum_{i=1}^{G} \mathcal{R}_{\text{seq}}(o_i), \quad \sigma^S = \sqrt{\frac{1}{G}\sum_{i=1}^{G} (\mathcal{R}_{\text{seq}}(o_i) - \mu^S)^2 + \epsilon}.
\end{gather}
}

where $\mu_j^Q$ and $\sigma_j^Q$ are the position-specific statistics for the $j$-th query slot while $\mu^S$ and $\sigma^S$ are the group statistics for the global sequence, $\lambda$ is a balancing coefficient and $\epsilon$ ensures numerical stability. This mechanism ensures that the policy simultaneously optimizes the specific utility of each query slot while satisfying global constraints, with both signals normalized to a consistent scale within the group.

\subsubsection{Continuous CTR Reward Adaptation}
To align query generation with evolving user interests, we incorporate a CTR reward derived from the production ranking model to provide supervision signals for the generated query. Since predicted probabilities are typically small and sparse, yielding negligible gradient signals, we apply linear scaling and upper-bound truncation. Finally, the CTR reward of $q_j$ for user $u$ is formulated as:

{\footnotesize
\begin{equation}
r_{\text{CTR}}(q_j) = \text{Clip}\left( \gamma \cdot \hat{p}_{\phi_T}(q_j \mid u), \, 0, \, \beta \right),
\label{eq:ctr_reward}
\end{equation}
}

where $\hat{p}_{\phi_T}$ denotes the predicted click probability from the CTR model with parameters $\phi_T$ on day $T$. $\gamma$ is a scaling factor to amplify reward variance for effective gradient estimation, and the clip operation prevents outlier scores from destabilizing the training process. This term $r_{\text{CTR}}(q_j)$ serves as a primary component of the composite query reward $\mathcal{R}_{\text{query}}(q_j)$ defined in Eq. (\ref{eq:reward_definitions}).

Note that the CTR model parameters $\phi_T$ are updated daily using the latest interaction logs $\mathcal{D}_T$ for tracking user interests and combating data drift, we also implement a daily continuous training mechanism to ensure the reward signal remains aligned with this up-to-date ranking system:

{\footnotesize
\begin{equation}
\phi_{T+1} \leftarrow \text{Train}(\phi_T, \mathcal{D}_T), \quad \pi_{\theta_{T+1}} \leftarrow \text{RL}(\pi_{\theta_T}, r_{\text{CTR}}^{\phi_{T+1}}).
\end{equation}
}

This iterative cycle ensures that the IL-GRPO framework continuously aligns the LLM's generation policy with the freshest user preferences, maintaining high recommendation utility over time.

\subsubsection{Decoupled Trigger Reward Optimization}

In \textsc{AIGQ-Think}, the model generates a set of trigger-query pairs $\{\mathcal{T}_k \rightarrow \mathcal{Q}_k\}_{k=1}^K$, where each trigger set $\mathcal{T}_k$ (a sequence of behavior indices) serves as an interpretable representation of a single user interest. To ensure clean credit assignment, we decouple the reward signals: the query side $\mathcal{Q}_k$ adopts the same reward formulation as in Section~\ref{subsubsec:query_reward}, while the trigger side $\mathcal{T}_k$ is supervised by a dedicated structural consistency objective.

Specifically, we also use ROUGE-L to define a trigger-level reward that measures the alignment between the generated triggers $\mathcal{T}_k^{\text{gen}}$ and the referenced label $\mathcal{T}_k^{\text{ref}}$:

{\footnotesize
\begin{equation}
r_{\text{trigger}}(\mathcal{T}_k) = \text{ROUGE-L}_F\big( \mathcal{T}_k^{\text{gen}}, \mathcal{T}_k^{\text{ref}} \big).
\label{eq:trigger_reward}
\end{equation}
}

To further prevent gradient interference between reasoning and recommendation, we enforce strict signal separation: tokens belonging to the trigger segment receive advantages derived solely from $r_{\text{trigger}}$, while query tokens use the composite advantage defined in Eq.~\eqref{eq:fine_grained_advantage}. Concretely, given $G$ sampled rollouts, we compute group statistics $\mu_k^{\mathcal{T}}$ and $\sigma_k^{\mathcal{T}}$ over the trigger rewards, and normalize the advantage for each trigger token $t$ as:

{\footnotesize
\begin{equation}
A_{i,t} = 
\frac{r_{\text{trigger}}(\mathcal{T}_{i,k}) - \mu_k^{\mathcal{T}}}{\sigma_k^{\mathcal{T}}}
\quad \text{if } t \in \text{tokens}(\mathcal{T}_{i,k}),
\label{eq:hybrid_advantage}
\end{equation}
}

This segmented advantage design ensures that the reasoning process focuses on understanding user interests while the query generation part optimizes for practical utility. As a result, the two components learn independently yet jointly improve overall recommendation quality.

\subsubsection{Entropy-Guided Training Stabilization} Motivated by recent work that leverages token-level entropy to stabilize policy updates\cite{wang2025stabilizingknowledgepromotingreasoning, wang20258020rulehighentropyminority, yu2025dapoopensourcellmreinforcement}, we design two entropy-based strategies to maintain exploration diversity and prevent policy collapse:

\begin{itemize}[left=0pt]
\item \textbf{Dynamic Entropy Regularization:} We add an entropy regularization term $\beta(s)\mathcal{H}(\pi_\theta)$ to the loss, where $\mathcal{H}(\pi_\theta)$ is the policy entropy and $\beta(s)$ is a linearly decaying weight that decreases with the global training step $s$. This encourages exploration early in training and reduces randomness as learning progresses.
\item \textbf{Entropy-Adaptive Clipping:} We apply differentiated clipping bounds for tokens with varying confidence levels as follows:

{\footnotesize
\begin{equation}
\epsilon(e_t) =
\begin{cases}
\epsilon_{\text{high}}, & \text{if } e_t \geq \tau, \\
\epsilon_{\text{low}},  & \text{if } e_t < \tau,
\end{cases}
\label{eq:entropy_clipping}
\end{equation}
}
\end{itemize}

where $e_t$ denotes the entropy of token $t$ and $\tau$ is the entropy threshold defined as the 80th percentile of token-level entropies in the current training batch. The bounds satisfy $\epsilon_{\text{high}} > \epsilon_{\text{low}}$, granting uncertain tokens more flexibility in policy updates while constraining confident tokens to maintain stability.

Beyond these, we further improve training stability with two complementary practices: (1) \textit{on-policy updates}~\cite{he2025skywork}, where rollouts are freshly sampled at each training step to ensure the behavior and target policies remain close in distribution; and (2) \textit{difficulty-based data filtering}, which focuses training on samples with intermediate reward scores to avoid trivially easy or highly noisy examples. 
Together, these techniques enhance training stability, encourage diverse reasoning paths and improve prediction accuracy in our list generation task for HintQ.

\section{Online Deployment}
To integrate a personalized LLM into a high-traffic industrial recommender under strict latency and resource constraints, we propose a hybrid retrieval architecture that combines: (1) \textsc{AIGQ}-u2q nearline recall, which runs \textsc{AIGQ-Direct} inference asynchronously and decoupled from user requests, then caches outputs for later use as personalized user-to-query (u2q) recommendations; (2) \textsc{AIGQ}-x2q real-time recall, which converts \textsc{AIGQ-Think}'s CoT outputs into a compact trigger-to-query (x2q) index for low-latency retrieval. Both components plug seamlessly into the existing multi-source recall and ranking stack of the online HintQ system.

\vspace{-0.2cm}
\subsection{AIGQ-u2q Nearline Recall}

Note that with a few hundred input tokens and dozens of output tokens, our production LLM service incurs several seconds per request, which far exceeds the strict sub-second budget of online latency. Therefore, we perform asynchronous inference and cache the personalized outputs for on-demand retrieval. Specifically, we develop a dedicated nearline module that monitors each user’s interaction stream and triggers an \textsc{AIGQ-Direct} inference upon detecting meaningful behavioral signals. Finally, the inference results are written to a user-level cache implemented as an internal key-value table, with each user retaining only the most recent $M$ inference runs. During online serving, the cached queries that reflect user's recent interest will be consumed as a recall source integrated into the current HintQ online stack. Under this approach, the best possible balance between deep personalization and strict online latency requirements is achieved, but meanwhile, it incurs a small degradation in personalization fidelity, since current signals are only reflected in subsequent visits.

\vspace{-0.2cm}
\subsection{AIGQ-x2q Real-time Recall}
\label{subsec:aigq-x2q}
Although \textsc{AIGQ-Think} improves personalized accuracy and discovery capability through CoT reasoning, its think-then-respond paradigm requires high inference latency. Meanwhile, the nearline \textsc{AIGQ-Direct} recall operates asynchronously, so when a request arrives, the user’s most recent intent such as the last clicked item may not yet be captured in the \textsc{AIGQ}-u2q cache. To bridge this recency gap, we distill the reasoning outputs of \textsc{AIGQ-Think} into a compact trigger-to-query (x2q) recall mechanism that supports real-time retrieval.

Specifically, the x2q index is constructed in an offline pipeline with daily update frequency. Given rich user behavioral traces as input, \textsc{AIGQ-Think} first generates CoT reasoning sequences of the form $\{\mathcal{T}_k \rightarrow \mathcal{Q}_k\}_{k=1}^K$. During index construction, each trigger $x_i \in \mathcal{T}_k$ is mapped to the full query set $\mathcal{Q}_k$, yielding entries $x_i \mapsto \mathcal{Q}_k$. We then rank the candidate queries associated with each trigger $x$ using a composite scoring function that balances semantic relevance and business effectiveness as follows:

{\footnotesize
\begin{equation}
S_{\mathrm{final}}(x, q) = \alpha \cdot S_{\mathrm{rel}}(x, q) + (1 - \alpha) \cdot S_{\mathrm{eff}}(q),
\end{equation}
}

where $S_{\mathrm{rel}}(x, q)$ is the relevance score reflecting how strongly the reasoning associates query $q$ with trigger $x$ while $S_{\mathrm{eff}}(q)$ is the effectiveness score aggregating multiple normalized business signals, such as page views and predicted CTR. The final score $S_{\mathrm{final}}(x, q)$ is a linear combination of these two scores with a tunable trade-off parameter $\alpha \in (0, 1)$.

\begin{table*}[t]
\centering
\caption{General performance on Taobao test set. The best results are illustrated in bold and the number underlined is the runner-up.}
\setlength{\abovecaptionskip}{-2em}
\setlength{\belowcaptionskip}{-1em}
\label{tab:offline}
\begin{tabular}{lcccc}
\toprule
Model & Cate HR@30 & Sem. Sim. & Query HR@30 & Unique Cates \\
\midrule
EBR  & 0.1998 & 0.5198 & 0.0100 & 8.4 \\
\midrule
Qwen3-30B-A3B &0.3054 & 0.5554 & 0.0022 & 5.4\\
Gemini 3 Pro &0.3449	&0.5475	& 0.0017& 7.0 \\
GPT-5.1 &0.3353	&0.5606	&0.0021	& 4.6 \\
\midrule
$\textsc{AIGQ-Direct}_{SFT}$& 0.4181 & 0.5926 & 0.0428 & 7.1 \\
$\textsc{AIGQ-Think}_{SFT}$ & 0.4437 & 0.6358 & 0.0549 & 7.7 \\
$\textsc{AIGQ-Direct}_{IL-SFT}$& 0.4305 & 0.5946 & 0.0442 & 7.5 \\
$\textsc{AIGQ-Think}_{IL-SFT}$ & 0.4653 & 0.6478 & 0.0559 & \textbf{10.3} \\

$\textsc{AIGQ-Direct}_{IL-SFT+GRPO}$ &0.3906 & 0.6116 & 0.0634 & 4.7 \\
$\textsc{AIGQ-Think}_{IL-SFT+GRPO}$ & 0.4438& \underline{0.6421} & 0.0614 & 8.0 \\
\midrule
$\textsc{AIGQ-Direct}_{IL-SFT+IL-GRPO}$ & \underline{0.4695} & 0.6341 & \underline{0.0679} & 6.7 \\
$\textsc{AIGQ-Think}_{IL-SFT+IL-GRPO}$ &\textbf{0.4704} & \textbf{0.6624} & \textbf{0.0745} & \underline{9.8} \\
\bottomrule
\end{tabular}
\end{table*}

\section{Experiments}
In this section, we present extensive offline experiments on real-world industrial datasets, together with rigorous online A/B testing, to validate the effectiveness and practicality of \textsc{AIGQ}. 

\subsection{Experimental Settings}
\subsubsection{Dataset.} Our dataset comprises highly reliable user-interaction pairs extracted from Taobao's online e-commerce logs spanning October 2025 to November 2025 with days 1–30 for training and days 31–32 for testing. After sampling, the dataset contains approximately 1 million page views (PVs), where training instances are organized at the PV level, while test instances are aggregated at the user level over a single day.

\subsubsection{Evaluation Metrics.} As a retrieval component, \textsc{AIGQ} is measured primarily by Query Hit Rate at K (Query HR@K) and Category Hit Rate at K (Cate HR@K). In order to verify the diversity impact, we supplement HR@K with Semantic Similarity (Sem. Sim) and the number of unique leaf categories (Unique Cates).

\subsubsection{Baseline Methods.}
We compare \textsc{AIGQ} against three major families of baselines: (i) traditional embedding-based retrieval (EBR), instantiated as a two-tower DNN model which is the strongest non-generative recall channel in our online system, delivering the highest PVR with competitive CTR; (ii) zero-shot LLMs, where we select state-of-the-art (SOTA) models spanning both open-source and proprietary ecosystems including Qwen3-30B-A3B, Gemini 3 Pro and GPT-5.1, all generating recommendations without any domain adaptation. (iii) task-adapted LLM variants under the \textsc{AIGQ} framework. Specifically, models with the subscript “SFT” are optimized via standard SFT lacking interest-aware modeling or prompt compression while those with “GRPO” apply the standard GRPO that computes advantages solely at the sequence level.

\subsubsection{Implementation Details}  
\textsc{AIGQ} is built upon Qwen3-30B-A3B with a 30B sparse Mixture-of-Experts (MoE) architecture and deployed on the PPU 810E platform. During training, the model is fine-tuned for two epochs using the Adam optimizer with a batch size of 1024 and an initial learning rate of $2 \times 10^{-3}$ in the Megatron framework~\cite{shoeybi2019megatron}.  In the IL-GRPO reinforcement learning phase, we use the ROLL framework~\cite{wang2025reinforcementlearningoptimizationlargescale} with a rollout batch size of 512 and a learning rate of \(1 \times 10^{-6}\). For each input sample, we generate 8 responses using a sampling temperature of 1.0 and top-\(k\) = 100. Difficulty-based data filtering is applied with reward thresholds set to the range \([0.1, 0.95]\). For stabilization, we clip the value estimates to \(\pm 0.5\), rewards to \(\pm 10\), and advantages to \(\pm 2.0\).


\subsection{Offline Evaluation}
Comprehensive experiments conducted on Taobao test set consistently show that AIGQ achieves superior performance compared to SOTA methods. As illustrated in Table~\ref{tab:offline}, the key findings are summarized as follows:

\begin{itemize}[leftmargin=*]
    \item \textbf{LLMs demonstrate significantly stronger potential for recommendation than traditional embedding-based methods.} Despite the lack of supervision signals from online logs, zero-shot LLMs consistently outperform EBR in category-level relevance and semantic alignment by a substantial margin of over 50\% in Cate HR@30, which is remarkable given that EBR is our strongest production recall channel. This highlights that LLMs, through pretraining on massive text, inherently capture semantic mappings between user intents and query categories, thus enabling effective query generation even without task-specific adaptation.

    \item \textbf{Adapting training paradigms to the list-wise nature of generative retrieval is essential.} Under the list-wise generation task, models trained with IL-SFT and IL-GRPO consistently outperform those using standard SFT or canonical GRPO. This is because IL-SFT explicitly captures ranking signals that reflect user interest density, while IL-GRPO enables fine-grained credit assignment by computing advantages at the query-level with diverse rewards to better align with real user feedback. Moreover, the combination of IL-SFT and IL-GRPO yields the best overall performance, demonstrating that list-aware supervised learning and fine-grained reward optimization are complementary.

    \item \textbf{Reasoning-augmented (\textsc{AIGQ-Think}) variants consistently surpass direct generation (\textsc{AIGQ-Direct}) across both accuracy and diversity.} This stems from their ability to unlock greater generative potential by structuring inference explicitly into two stages: first distilling user intent into coherent interest points, then generating queries grounded in this structured representation. This separation empowers the model to simultaneously achieve higher precision and broader coverage than direct implicit generation.
\end{itemize}

\subsection{Ablation Study}


\begin{table}
\centering
\caption{The experimental results of ablation study.}
\setlength{\abovecaptionskip}{-2em}
\setlength{\belowcaptionskip}{-1em}
\label{tab:Ablation2}
\begin{tabular}{lcc}
\toprule
Model & Cate HR & Query HR  \\
\midrule
$\textsc{AIGQ-Direct}_{SFT}$& 0.4181  & 0.0428  \\
+ Interest-Guided Label Re-Ranking & 0.4260  & 0.0442  \\
+ Prompt Compression & 0.4305 & 0.0442 \\
+ IL-GRPO & 0.4695 & 0.0679  \\
\midrule
$\textsc{AIGQ-Think}_{SFT}$ & 0.4437  & 0.0549 \\
+ Interest-Guided Label Re-Ranking & 0.4492  & 0.0594  \\
+ Reasoning Augmentation & 0.4653 & 0.0559 \\
+ IL-GRPO & 0.4704 & 0.0745 \\
\bottomrule
\end{tabular}
\end{table}

We conduct an ablation study on the Taobao test set to assess the contribution of each component in \textsc{AIGQ} (Table~\ref{tab:Ablation2}). Note that due to space limitations, we only show the results on two representative metrics, Cate HR@30 and Query HR@30. Table~\ref{tab:Ablation2} reveals the following key insights:
\begin{itemize}[leftmargin=*]
    \item Interest-guided label re-ranking, which is a core part of IL-SFT, provides consistent though modest gains over the base SFT model by exposing the generator to supervision signals that reflect the underlying interest density of different user intents.

    \item Prompt compression preserves or slightly enhances performance when properly trained, indicating that eliminating redundant tokens does not degrade semantic quality, whereas reasoning augmentation yields a more substantial improvement by explicitly modeling user intent structure and guiding query generation through structured planning.

    \item IL-GRPO further delivers significant advancement after fine-tuning the model with IL-SFT, producing a large relative increase in both category-level and query-level precision. This confirms that by estimating advantages at the query level, IL-GRPO enables fine-grained credit assignment that better aligns RL with the true objective of top-$K$ generative retrieval.
\end{itemize}

\begin{table}[t]
\centering
\caption{Online performance of \textsc{AIGQ} across effectiveness, engagement, and discovery. 
Effectiveness and engagement report relative gains while discovery shows the percentage of queries or leaf categories uniquely covered by \textsc{AIGQ}.}
\setlength{\abovecaptionskip}{-2em}
\setlength{\belowcaptionskip}{-2em}
\begin{tabular}{l l c}
\toprule
\textbf{Category} & \textbf{Metric} & \textbf{Value (\%)} \\
\midrule
\multirow{3}{*}{Effectiveness} 
    & HintQ UCTR      & +7.42 \\
    & Attributed orders          & +10.31 \\
    & Attributed GMV             & +10.68 \\
\midrule
\multirow{2}{*}{Engagement} 
    & LT-7 retention  & +3.73 \\
    & Search UV       & +0.20 \\
\midrule
\multirow{2}{*}{Discovery} 
    & Unique queries            & 79.3 \\
    & Unique leaf categories    & 37.3 \\
\bottomrule
\end{tabular}
\label{tab:aigq_online_results}
\end{table}


\subsection{Online A/B Testing}
We conduct online A/B testing in the HintQ scenario of Taobao’s homepage over a thirty-day period, with results shown in Table~\ref{tab:aigq_online_results}. \textsc{AIGQ} consistently improves performance across effectiveness, engagement, and discovery. Notably, it achieves significant gains in core conversion metrics (e.g., +10.31\% in attributed orders and +10.68\% in GMV), indicating that its generated queries better capture nuanced user intent and enable more precise demand fulfillment. This enhanced relevance also drives stronger long-term engagement, reflected in a +3.73\% increase in 7-day retention, suggesting an effective balance between personalization and serendipity. Moreover, almost 40\% of leaf categories from \textsc{AIGQ} are not covered by any other recall channel, demonstrating strong diversity and differentiation.

\begin{figure}[htbp]
    \centering
    \setlength{\abovecaptionskip}{-0.1em}
    \setlength{\belowcaptionskip}{-2em}
    \includegraphics[width=0.9\linewidth]{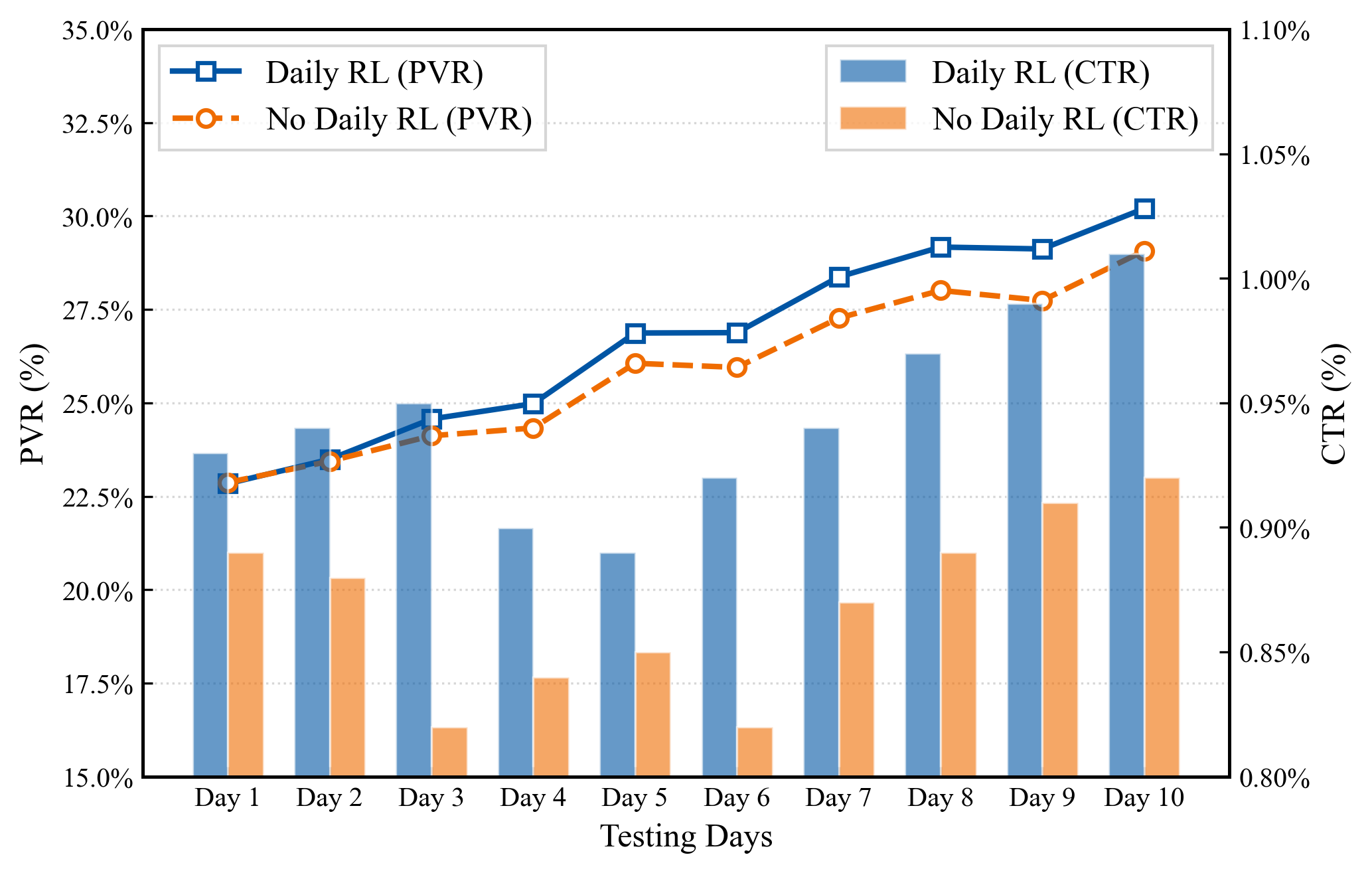}
    \caption{Online A/B test: Impact of daily RL updates.}
    \label{fig:rl_daily}
\end{figure}

Additionally, we evaluate the impact of daily policy updates in IL-GRPO via a ten-day A/B test. As shown in Figure~\ref{fig:rl_daily}, enabling daily reinforcement learning updates consistently improves both PVR and CTR. This confirms that continuous alignment with real-time user feedback through CTR-based rewards mitigates distribution shift between training and deployment. The result highlights a key insight for applying LLMs in recommendation: effective generative retrieval requires not only strong language understanding but also tight integration with live user signals and downstream ranking systems to ensure end-to-end alignment.

\section{Conclusion and Future Work}
In this work, we introduced \textsc{AIGQ}, the first end-to-end generative framework for personalized pre-search query recommendation. By synergizing direct and reasoning-augmented generation, list-wise reinforcement learning grounded in online CTR signals, and a hybrid offline-online deployment strategy, \textsc{AIGQ} achieves substantial gains across effectiveness, discovery, and business outcomes at billion-user scale.  Looking ahead, we envision two key directions to further advance this paradigm: (i) enriching query generation through more precise multimodal context understanding; (ii) enabling scalable real-time reasoning through efficient deployment techniques such as distillation or quantization.

\begin{acks}
We thank the Taobao Search and Recommendation teams for their support. This work was conducted at Alibaba Group.
\end{acks}

\bibliographystyle{ACM-Reference-Format}
\bibliography{references}  

\end{document}